\documentclass[twocolumn,secnumarabic,amssymb, nobibnotes, aps, prd]{revtex4-1}
\usepackage{amsmath}
\usepackage{graphicx}
\RequirePackage[pagewise,mathlines]{lineno}

\begin{document}
\title{Comment on ``$\eta_c$ production in photon-induced interactions at the LHC''}%

\author{Spencer R. Klein}%
\email[]{srklein@lbl.gov}
\affiliation{Nuclear Science Division, Lawrence Berkeley National Laboratory, Berkeley, CA 94720 USA}
\date{\today}%
\begin{abstract}

In ``$\eta_c$ production in photon-induced interactions at the LHC," \cite{Goncalves:2018yxc} Goncalves and Moreira discuss inclusive and exclusive  $\eta_c$ production at $pp$ and $pA$ collisions at LHC energies.   The exclusive channels are via two-photon and photon-Odderon interactions.  This comment points out that there is a large additional source of almost-exclusive $\eta_c$ in ultra-peripheral collisions: from the radiative decay of $J/\psi$ that are produced in photon-nucleon interactions.   Although the $J/\psi\rightarrow\gamma\eta_c$ branching ratio is small, the $J/\psi$ production cross-section is large enough that it dominates over the exclusive channels considered in \cite{Goncalves:2018yxc}, and is comparable to the non-exclusive production.  In $J/\psi\rightarrow\gamma\eta_c$, the photon is very soft and therefore easy to miss, and the $\eta_c$ will have very similar kinematics to the $J/\psi$.
\end{abstract}
\maketitle

Two-photon production of the $\eta_c$ has long been a target for ultra-peripheral collisions at heavy ion colliders \cite{Baron:1993az,Vidovic:1995bj,Schramm:1995ck}, and early RHIC  \cite{Nystrand:1998hw} and ALICE \cite{Sergey} studies considered its production in the two-photon channel.    Production via double-Pomeron (also called Central Exclusive Production) has also been considered \cite{Schramm:1996aa}.  These calculations found rather small cross-sections, leading to a decline in interest. 

Goncalves and Moreira \cite{Goncalves:2018yxc} consider $pp$ and $pA$ collisions,  via both two-photon and photo-nuclear channels, finding that inclusive $\gamma p$ interactions dominate, with cross-sections of 3.492 nb in $pp$ collisions at $\sqrt{s}=13$ TeV, and 3.194 $\mu$b in $pPb$ collisions at $\sqrt{s}=8.1$ TeV.  The total inclusive rates (two-photon + photon-Odderon \cite{Goncalves:2012cy}) are much smaller, 0.059 nb and 0.501 nb respectively for $pp$ collisions, and 0.182 $\mu b$ and 0.351 $\mu$b for $pPb$ collisions.   The existence of the Odderon is rather speculative, so it is important to consider whether there are any exclusive production channels channels with cross-sections larger than the $\gamma\gamma$ channel, to avoid possible false claims for the existence of the Odderon.

As Section 4.4 of Ref. \cite{Bertulani:2005ru} pointed out, the rate for coherent $J/\psi$ photoproduction, followed by the decay $J\psi\rightarrow\gamma\eta_c$ is larger than that for two-photon production of the $\eta_c$.   Because the emitted photon is so soft (energy of 111 MeV in the $J/\psi$ rest frame), it is likely to be missed in any LHC detector, and  the $\eta_c$ will have a similar rapidity and transverse momentum as the $J/\psi$ parent, so the final state will look like exclusive photoproduction.

Here, I use STARlight \cite{Klein:2016yzr} to make a similar calculation for the $pp$ and $pA$ collisions discussed in Ref.  \cite{Goncalves:2018yxc}.  STARlight uses a fairly standard model for photo-production \cite{Klein:1999qj}, using the Weizsacker-Williams photon flux and requiring that the two nuclei do not interact hadronically.    For $pp$ collisions, STARlight uses a photon flux \cite{Klein:2003vd} similar to that used by Goncalves and Moreira.  STARlight does not include nuclear shadowing, but, since the bulk of the production in $pA$ collisions comes when the proton is a target, shadowing should not be important here. 

Table \ref{tab:tab} gives the coherent photoproduction cross-sections for $J/\psi$ and $\eta_c$ production cross-sections for these $pp$ and $pPb$ collisions, based on the branching ratio Br($\eta_c\rightarrow\gamma\eta_c$)=0.17 \cite{Patrignani:2016xqp}.    
 \begin{table}[h]
 \begin{tabular}{|c|c|c|}
 \hline
   &  $\sigma(J/\psi)$ & $\sigma(J/\psi\rightarrow\eta_c\gamma)$  \\
   \hline
   $pp$ & 79 nb & 1.34 nb \\
   $pPb$ (proton-shine) & 3.21 $\mu$b & 54 nb  \\
   $pPb$ (lead-shine) & 57.5 $\mu$b & 1.0 $\mu$b \\
   \hline
\end{tabular}
\caption{The cross-sections for $J/\psi$ photoproduction in $pp$ and $pPb$ collisions at the LHC (with separate lines for photon emission by the proton and the lead nucleus), along with the production cross-section for the $\eta_c$ in all three modes.}
\label{tab:tab}
\end{table}

For both $pp$ and $pPb$ collisions, the $\eta_c$ rate is higher than the exclusive channels considered in Goncalves and Moreira.  The cross-sections are also higher than the non-exclusive diffractive production channel (which also leaves both beam particles intact).   
   
\begin{figure}
  \centering
  \includegraphics[width=0.9\linewidth]{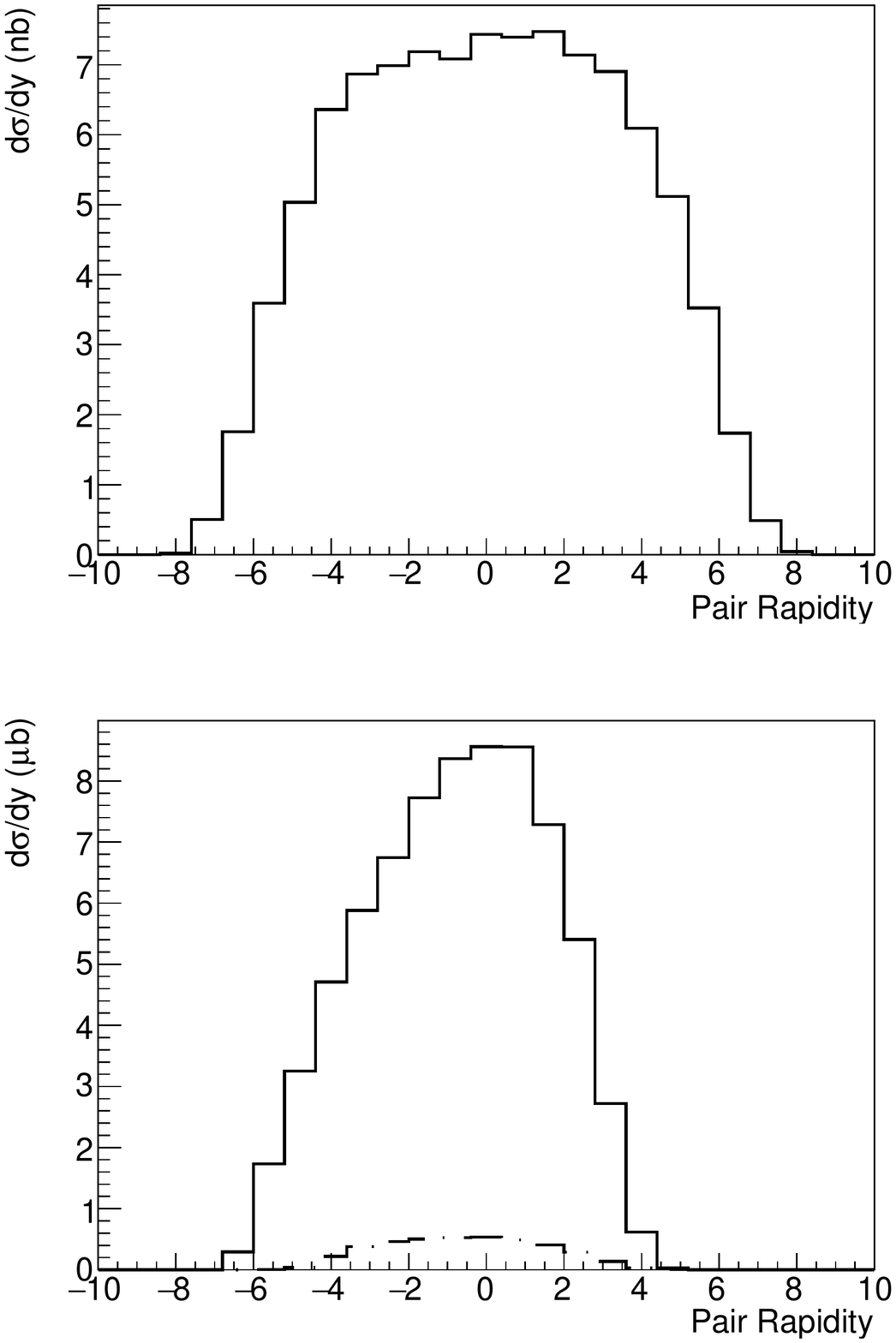}
  \caption{The rapidity distribution $d\sigma/dy$ for $J/\psi$ photoproduction at the LHC for (top) $pp$ collisions at $\sqrt{s}=13$  TeV and (bottom) $pPb$ collisions at $\sqrt{s_{NN}}=8.1$  TeV.  In the bottom plot, the solid black curve is 'lead-shine,' and the dashed dark blue curve is 'proton-shine.'}
  \label{fig:rapidity}
\end{figure}

Figure \ref{fig:rapidity} shows the rapidity distribution for $J/\psi$ produced in $pp$ and $pA$ collisions; the latter is separated out by photon emitter.  The bulk of the $pPb$ production occurs when photons are emitted by the lead nucleus.   Both overall distributions are
quite similar to the $d\sigma/dy$ for diffractive photoproduction in Fig. 4 of Ref. \cite{Goncalves:2018yxc}.

One way to differentiate between $\gamma\gamma$ and photon-Pomeron (or photon-Odderon) production is by examining the transverse momentum distribution, $p_T$, of the $\eta_c$.  Two-photon production has a considerably smaller $p_T$ scale than the photoproduction channels \cite{Baltz:2009jk}, particularly when a proton is the photoproduction target.  Photon-Pomeron and photon-Odderon processes should have a similar $p_T$ spectra, so be indistinguishable. 

These cross-sections are considerably larger than any of the exclusive $\eta_c$ cross-sections discussed in Ref. \cite{Goncalves:2018yxc}, and are the most likely source for apparent exclusive $\eta_c$.  Similar conclusions will apply for collisions of other beam particles \ \cite{Goncalves:2012cy}.  It is important that $J/\psi$ photoproduction be considered; otherwise, any observation of exclusive $\eta_c$ beyond that expected from two-photon physics might be mistaken for a more exotic process.    

This work was funded by the U.S. DOE under contract number DE-AC02- 05-CH11231.

\end{document}